\documentclass[global,twocolumn]{svjour}
\usepackage{url,amssymb,graphicx}
\begin{document}
\newcommand{\beq}{\begin{equation}}
\newcommand{\eeq}{\end{equation}}
\newcommand{\beqa}{\begin{eqnarray}}
\newcommand{\eeqa}{\end{eqnarray}}
\newcommand{\capsub}[2]{{#1}_{\scriptscriptstyle {\rm #2}}}
\newcommand{\smsub}[2]{{#1}_{\rm #2}}
\newcommand{\Lya}{Ly$\alpha\,$}
\newcommand{\Omat}{\capsub{\Omega}{M,0}}
\newcommand{\Olam}{\capsub{\Omega}{\Lambda,0}}
\newcommand{\Otot}{\capsub{\Omega}{T,0}}
\newcommand{\fth}{\smsub{f}{th}}
\newcommand{\fobs}{\smsub{f}{obs}}
\newcommand{\Fth}{\smsub{\tilde{f}}{th}}
\newcommand{\Fobs}{\smsub{\tilde{f}}{obs}}
\newcommand{\Obar}{\smsub{\Omega}{bar}}
\newcommand{\lPl}{\capsub{\ell}{Pl}}
\newcommand{\rPl}{\capsub{\rho}{Pl}}
\newcommand{\rv}{\smsub{\rho}{v}}
\newcommand{\Rmin}{\smsub{R}{min}}

\title{Wolfgang Priester: from the big bounce to the $\Lambda$-dominated universe}

\dedication{In memoriam Wolfgang Priester, 22 April 1924 -- 9 July 2005}

\author{James Overduin\inst{1}
\and Hans-Joachim Blome\inst{2}
\and Josef Hoell\inst{3}}
\institute{Gravity Probe B, Hansen Experimental Physics Laboratory,
   Stanford University, Stanford, CA 94305-4085, U.S.A.;
   \email{overduin@relgyro.stanford.edu}
\and University of Applied Sciences, Aerospace Department, Fachhochschule
   Aachen, D-52064 Aachen, Germany;
   \email{blome@fh-aachen.de}
\and German Aerospace Center (DLR), Space Management, Space Science
   (RD-RX), D-53227 Bonn, Germany;
   \email{josef.hoell@dlr.de}}
\maketitle

\abstract{Wolfgang Priester was one of Germany's most versatile and quixotic
astrophysicists, re-inventing himself successively as a radio astronomer,
space physicist and cosmologist, and making a lasting impact on each field.
We focus in this personal account on his contributions to cosmology, where
he will be most remembered for his association with quasars, his promotion
of the idea of a nonsingular ``big bounce'' at the beginning of the current
expansionary phase, and his recognition of the importance of dark energy
(Einstein's cosmological constant $\Lambda$) well before this became the
standard paradigm in cosmology.
\keywords{cosmology: big bang -- dark energy -- cosmological parameters -- quasars: absorption lines}}

\section{Early Career(s)}
Born in Detmold in 1924, Wolfgang Priester began his university studies in
astronomy, mathematics and physics at G\"ottingen in 1946.  One of his
teachers was Theodor Kaluza, the father of modern higher-dimensional
unified field theory, whom Wolf remembered as ``not very demanding.''
He obtained his doctorate in 1953 with a thesis on photometry with the
sodium D~line [1-3]. 
His subsequent scientific career can be divided into four distinct phases.
The first of these, devoted to radio astronomy and astrophysics, began
immediately with two years under Albrecht Uns\"old in Kiel [4-8]
and continued in Bonn, where Wolf helped in the construction of the 25~m
Stockert radio telescope and completed a major survey of the radio sky at
200~MHz with Franz Dr\"oge in 1956 \cite{wolf56}.  This period saw the
publication of his first article in {\it Naturwissenschaften} (on radio
emission from comets \cite{wolf57b}), as well as an influential 1958 study
of the statistics of extragalactic radio sources (based on his
{\em Habilitation\/} thesis under Friedrich Becker and cosmologist
Otto Heckmann \cite{wolf58b}) which played a role in the then-raging conflict
between big-bang and steady-state theories \cite{Kra96a}.  There were other
papers during this period [12-14], 
but Wolf's greatest legacy as a radio astronomer was probably a practical one:
as founding director of the Institute for Astrophysics and Space Research at
the University of Bonn (1964-89),  he initiated (together with Friedrich
Becker and Otto Hachenberg) the construction of the world's largest fully
steerable radio telescope at Effelsberg.  The performance of this 100~m
instrument, which was inaugurated in 1971, remained unsurpassed for thirty
years.

Wolfgang Priester's second scientific career, that as a space physicist, began
when he used radio transmissions from the newly launched {\em Sputnik}
satellite to compute its orbit in 1958 \cite{wolf58c,wolf58e}, a feat which
resulted in his invitation to join NASA's Goddard Institute for Space
Studies in New York City between 1961 and 1964.  He was among the first to use
such signals to derive physical models of the Earth's atmosphere, particularly
as it is affected by solar activity.  This work that led to a second article
in {\it Naturwissenschaften} \cite{wolf59c} as well as many other papers
[18-37], 
including four in {\it Nature}.  Of these publications, and Wolf's guiding
roles in bodies such as COSPAR (the Committee for Space Research) during
these years, probably no contribution has proved more influential than a
detailed model of the upper atmosphere that he developed together with
Isadore Harris during his years in the U.S.A. [38-47]. 
The Harris-Priester model remains in use by NASA and is still taught in
graduate schools after more than forty years.

The third subject that exerted a lifelong fascination on Wolfgang Priester
can also be traced back to his time in the U.S.A.  Wolf was present at the
epochal First Texas Symposium on Relativistic Astrophysics in Dallas in
December of 1963, where the problem of the immense energy output of the
objects then known as ``quasi-stellar radio sources'' was first confronted
in earnest.  This was the meeting at which Thomas Gold famously speculated
``that the relativists with their sophisticated work were not only
magnificent cultural ornaments but might actually be useful to
science'' \cite{Gol65}.

Less well known is how these enigmatic objects eventually became known as
``quasars.''  The full story has been told by Nigel Calder, who notes
ironically that it took native German and Chinese speakers to come up with
a decent acronym during meetings at the Goddard Institute in New York City
\cite{Cal03a}.  Wolf Priester's initial suggestion of ``quastar'' was
rejected by his colleague Hong-Yee Chiu on the grounds that it sounded too
much like the name of the telescope maker ``Questar.''  Chiu shortened
Wolf's proposal to ``quasar'' \cite{Chi64}, the new term was picked up by the
{\it New York Times} beginning in April 1964 (and reluctantly accepted by the
{\it Astrophysical Journal\/} in 1970 \cite{Sch70}), and that is how the most
luminous objects in the Universe got their name.  Quasars remained a source
of never-ending wonder for Wolf, who co-authored a major review of their
properties with Johan Rosenberg in 1965 \cite{wolf65b}, later expanded in 1968
\cite{wolf68a}.  He returned to the subject in publications throughout his
tenure as director of the Institute of Astrophysics and Space Research in Bonn
[50-59], 
and hosted guests such as Thomas Gold, Bob Jastrow and Maurice Shapiro over
the years.  It is likely that these relativistic engines at the edge of the
universe also drew him back to his first and last love: cosmology.

\section{Cosmology and the vacuum}

Wolfgang Priester's fourth career, that as a cosmologist, occupied most of his
last two decades and accounted for almost half of his more than 100
publications.  The beginning of his relationship with this field can however
already be seen in his second paper on radio sources and the value of Hubble's
constant in 1954 \cite{wolf54a}, and in his 1958 treatment of the statistics
of extragalalactic radio sources in relativistic cosmology \cite{wolf58b}.
His return to the field dates from a 1982 survey \cite{wolf82b} and two 1983
publications titled ``From the big bang to black holes'' \cite{wolf83a} and
``Where is the antimatter?'' \cite{wolf83b}.  These were followed by a 1984
address on cosmic evolution \cite{wolf84a} and a series of review articles with
Hans-Joachim Blome in {\it Naturwissenschaften} covering almost all aspects of
what is now called physical cosmology [64-66]. 
When these articles appeared, this field was still in its infancy; the
definitive text (by James Peebles) had only been circulating for a few years
and the particle-cosmology revolution heralded by such authors as Rocky Kolb
and Michael Turner was five years away.  For many, the word ``cosmology''
still conjured up something close to metaphysics.  Wolf Priester was in the
vanguard as it grew into a full-fledged empirical science.

\begin{figure}
\resizebox{\hsize}{!}
{\includegraphics{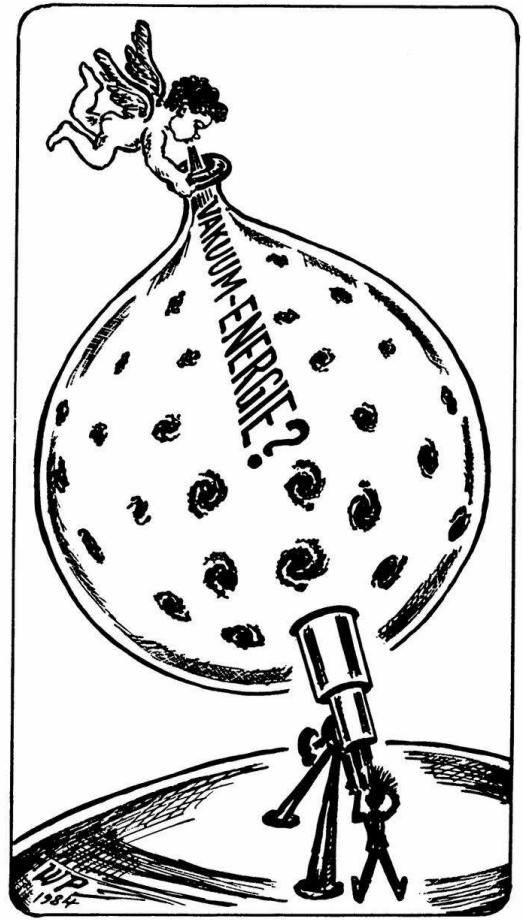}}
\caption{{\em Vacuum energy\/}, by Wolfgang Priester (1984).}
\label{fig1}
\end{figure}

At the heart of these articles and later ones with Blome and Josef Hoell was
Wolf's insistence on confronting the physical reality of vacuum energy
(represented in Einstein's theory of general relativity by the symbol
$\Lambda$, and now more popularly known as ``dark energy'').
This insistence made him something of a prophet in the wilderness.
Most cosmologists at the time preferred to ignore the $\Lambda$-term
on the grounds that no value satisfying observational bounds could be
understood theoretically, while no reasonable theoretical expectations for
this quantity made sense observationally (i.e., the cosmological constant
problem, which persists today).  Wolf was fond of emphasizing that
Einstein himself had recognized the importance of this issue, referring to
vacuum energy in 1920 as the ``new ether of general relativity'' \cite{wolf84d}.
His characteristically playful 1984 sketch captured the perplexity
of the astronomer in the face of this ``new ether'' (Fig.~\ref{fig1}).

\section{The big bounce}

From the physical reality of dark energy, together with the fact that upper
limits on its energy density are many orders of magnitude smaller than one
expects based on quantum field theory, Blome and Priester were led in 1984
to the possibility of vacuum {\em decay\/} and its corollary, a
time-dependent cosmological ``constant'' based, for example, on the vacuum
expectation value of a scalar field \cite{wolf84d,wolf85b}.  Largely taboo
at the time, these subjects have since grown into an industry under the
name of ``quintessence'' \cite{Cal98}\footnote{Caldwell et al were not the
   first contemporary scientists to revive this venerable term, as is often
   supposed: it was formally introduced into geology and cosmochemistry during
   NASA's {\it Apollo\/} program \cite{Pap71,Lew75}.  We thank an anonymous
referee for pointing this out.} although the underlying idea is now often
attributed to Christoph Wetterich and others in 1988 \cite{Wet88,Pee88}.
We would argue that it goes back at least to work carried out by Ernst
Streeruwitz in 1975 \cite{Str75} and cited in Refs.~\cite{wolf84d,wolf85b}.

After several more projects [68-71], 
including an elegant de-mystification of cosmological recession velocities
in {\it Naturwissenschaften} \cite{wolf87b}, Wolfgang Priester returned to
the $\Lambda$ term in 1987.  This time he set his sights on the vacuum's
potential for addressing the problem of the initial singularity --- an issue
which was never far from his mind, and in which we can probably discern the
early influence of Otto Heckmann.  Wolf and his colleagues were among the
first to take seriously the possibility that a phase transition might separate
the early universe from a primordial de~Sitter-like era, so that the contents
of the present universe might literally have sprung from the pure vacuum
energy of that earlier epoch.  A de~Sitter phase further opened up the
possibility that the expanding universe began in a nonsingular ``big bounce''
rather than a big bang singularity [73-75]. 
Wolf raised an early question as to whether vacuum energy or relativistic
particles would dominate in a pre-inflationary universe, a debate that
continues today.

\begin{figure}
\resizebox{\hsize}{!}
{\includegraphics{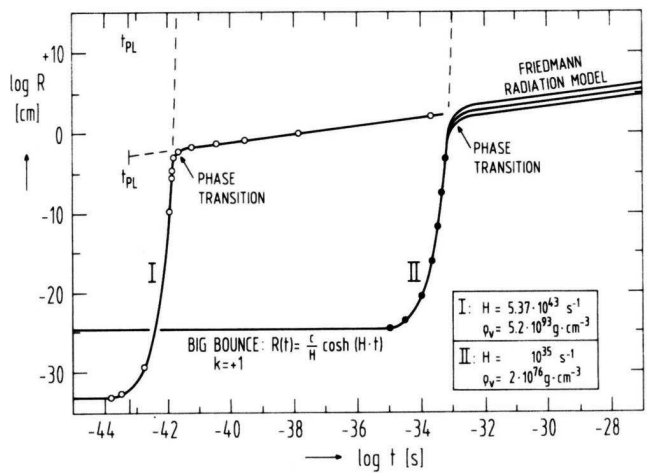}}
\caption{Two possible scenarios by which a ``big bounce'' in the very early
   universe joins smoothly to the standard cosmological model.
   Model~I (due to Mark Israelit and Nathan Rosen \cite{Isr89}) is
   characterized by Planck size $\lPl$ and density $\rPl$ at the bounce
   (typical of models within quantum cosmology), while Model~II has
   $\rv\approx10^{-18}\rPl$, corresponding to an energy of $10^{14}$~GeV
   necessary for the production of X-bosons which are assumed to transform
   into quarks and leptons at the end of inflation \cite{wolf91b}.}
\label{fig2}
\end{figure}

It proved possible to be quite specific about the properties that such a
bounce must have, if it is to satisfy natural initial conditions and also
join smoothly to standard radiation-dominated cosmology at later times.  
Fig.~\ref{fig2} illustrates the range of allowed models for pre-inflationary
scenarios dominated by vacuum energy.
The energy density $\rv$ of the primordial quantum vacuum lies in the range
$10^{-18}\rPl\leqslant\rv\leqslant\rPl$, where $\rPl$ is the Planck density,
and the curvature radius $\Rmin$ at the bounce can be shown to satisfy
\cite{wolf91b}
\beq
\lPl \leqslant \Rmin = \sqrt{3c^2/(8\pi G\rv)} \leqslant 2\times10^8\lPl \; ,
\eeq
where $\lPl$ is the Planck length.

Qualitatively, of course, bounce models have a venerable history and were
endorsed on largely philosophical grounds by such thinkers as Willem de~Sitter,
Carl Friedrich von Weizs\"acker, George McVittie and George Gamow (who
however stressed that ``from the physical point of view we must forget
entirely about the pre-collapse period'') \cite{Kra96b}.  The advent of 
inflation in the 1980s gave them a new impetus \cite{Pet82}, and they remain
in vogue today among inflationary model-builders \cite{Gra03}, string
theorists \cite{Gas03} and quantum cosmologists \cite{Boj05}.  With all this
attention, it is worth recording that the term ``big bounce'' appears
to have been introduced to cosmology by Wolfgang Priester
in 1987.\footnote{It was re-introduced the following year in Iosif Rozental's
   {\it Big Bang, Big Bounce}, a revised translation of a Russian book (by a
   different title).
   The phrase apparently first appeared as the title of a 1969 novel
   by Elmore Leonard.}

\section{The $\Lambda$-dominated universe}

\begin{figure}
\resizebox{\hsize}{!}
{\includegraphics{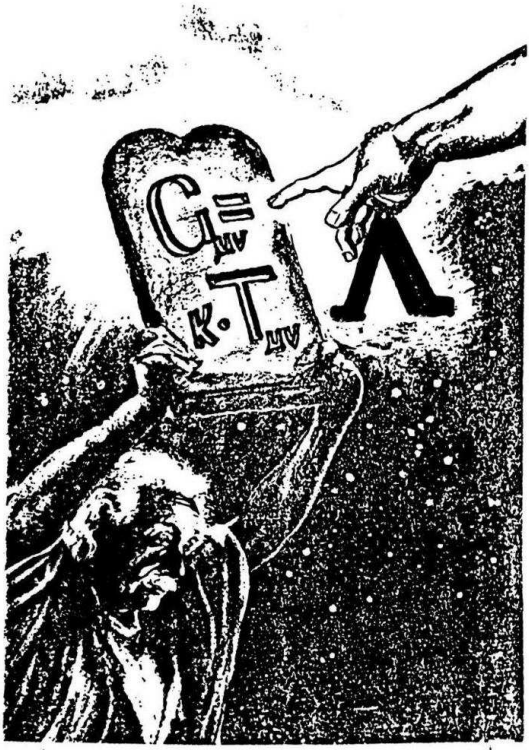}}
\caption{{\em The Ten Complete Commandments of General Relativity\/},
   adapted by Wolfgang Priester (1992) from the woodcut by Gustav Dor\'e}
\label{fig3}
\end{figure}

By 1988, Wolf had concluded that ``the {\em a priori\/} restriction to
models with $\Lambda=0$ is not justified'' in physical cosmology
\cite{wolf88a,wolf88b}, a view he urged on readers with another of his
wonderful drawings (Fig.~\ref{fig3}).
The basis for this conclusion was both observational and theoretical.
A universe with $\Lambda=0$ could not have lasted as long as the oldest stars
unless it were now expanding very slowly, in contradiction with most
observational data.  This ``age problem'' became particularly acute in
the newly popular inflationary paradigm, which implied the existence of
large amounts of gravitating dark matter, thereby decelerating the Universe
even more quickly.  If, instead of weighing down the Universe with more
matter, one filled it with self-repelling dark {\em energy\/}, deceleration
would be counteracted and the problem solved.  Thus, after intervening work
on baryon asymmetry and other topics
\cite{wolf88c,wolf88d,wolf88e,wolf89a,wolf89b,wolf90a,wolf90b},
Wolf wrote with Josef Hoell in 1991 that ``if the Hubble constant is greater
than 50 km s$^{-1}$ Mpc$^{-1}$ ... the cosmological constant {\em must be
greater than zero\/} ... The apparently missing fraction of 90\% or more of
cosmic matter is entirely compensated for by the cosmological constant''
\cite{wolf91a}.  He was among the very first to draw this conclusion
\cite{Efs90}, and did so four years before some of the more celebrated names
that have become attached to it since \cite{OS95,KT95}.

Wolf also recognized a more aesthetic motivation for a nonzero $\Lambda$
term, based primarily on its role as a fundamental length scale in a
(closed) universe via the de~Sitter radius $R_{\Lambda}=1/\sqrt{\Lambda}$
\cite{wolf95e}.  Those who had the pleasure of discussing cosmology
with him for any length of time were likely to become familiar with the
late-time limit of the Friedmann-Lema\^{\i}tre equation:
\begin{equation}
\Lambda c^2 = 3 H_{\infty}^2 \; .
\end{equation}
The deceptive simplicity of this relationship between a dynamical quantity
(the expansion rate $H$) and an apparently fundamental constant of nature
($\Lambda$) was a source of endless fascination to Wolf, so much so that he
joked that he would like to have it engraved on his tombstone.  In this he
would have found a fellow believer in Arthur Eddington, who wrote that there
must be a cosmological constant because ``an electron could never have decided
how large it ought to be, unless there existed some length independent of
itself for it to compare itself with'' \cite{Edd24}.

\section{The Bonn-Potsdam model}

\begin{figure}
\resizebox{\hsize}{!}
{\includegraphics{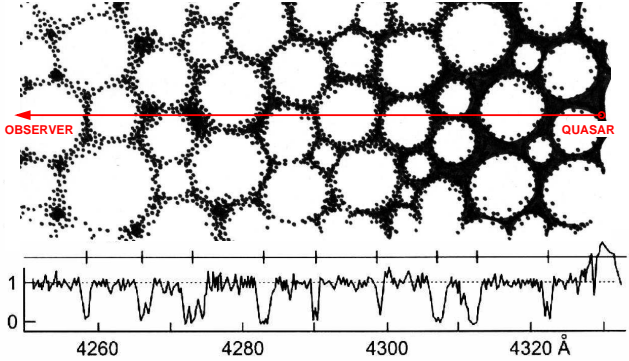}}
\caption{The sponge-like distribution of absorbing material intersecting the
   line of sight to a distant quasar and producing the characteristic \Lya\
   absorption spectrum (bottom).  The characteristic scale of the absorber
   distribution is approximately constant in comoving coordinates.
   (Adapted from a drawing by W.~Priester in 1995 \cite{wolf95a})}
\label{fig4}
\end{figure}

For Wolfgang Priester, final confirmation of the existence of dark energy
appeared courtesy of his old friends, the quasars.  Spectra taken in the
direction of these distant objects show characteristic absorption lines making
up what is known as the Lyman-$\alpha$ (\Lya) forest.  These lines arise
when the light from the quasars passes through intervening concentrations of
gas which are thought to be distributed around emptier regions known as voids
(Fig.~\ref{fig4}).  With Josef Hoell, he realized in 1991 that these lines
could be used as tracers of cosmic expansion, if intrinsic evolution in the
absorber population could be neglected in comparison with the Hubble expansion
rate \cite{wolf91c}.  Together with Dierck-Ekkehard Liebscher, Priester and
Hoell applied this method to a set of high-resolution spectra from 21 quasars
between 1992 and 1994 and found that the spacings of the absorption lines
varied with redshift in a way that, if attributed entirely to cosmic expansion,
implied a spatially closed, $\Lambda$-dominated world model whose expansion
was {\em accelerating\/}, not decelerating as was almost universally believed
at the time.  Their best-fit model, which came to be known as the
``Bonn-Potsdam'' or BN-P model, was characterized by a dark-energy density of
$\Olam\approx1.08$ and a total matter density of $\Omat\approx0.014$,
as measured in units of the critical density [88-93].

The BN-P model attracted little attention, its large value of $\Olam$ in
particular ``falling outside the range of theories considered polite in the
mid-1990s,'' as Wolf later commented \cite{Cal03b}.  He was fond of recounting
a story from the 1994 Erice school on high-energy astrophysics, where one
well-known cosmologist told him, ``Your $\Lambda$ is outrageous!  It does not
exist --- it is zero!''  Four years later, the cosmological tide turned with
the announcement by two teams of supernova observers in the U.S.A. that $\Olam$
{\em was\/} in fact greater than zero, and probably greater than $\Omat$,
implying that we live in an accelerating universe --- developments that were
hailed by {\it Science} magazine as the Breakthrough of the Year in 1998
\cite{Gla98}.  The same cosmologist wrote to Wolf on his birthday that year:
``Well, there are still new things to learn.'' ``Saulus became Paulus,''
Wolf always concluded with satisfaction in his talks.

His early measurement of $\Olam$ using quasar absorption lines is recognized
as follows in the {\it Oxford Guide to Modern Science}: ``Historians of
science may note that Wolfgang Priester of Bonn and Dierck-Ekkehard Liebscher
of Potsdam reported the cosmic acceleration in 1994.  That was more than three
years before other astronomers, with great fanfare, announced the acceleration
seen by observing exploding stars'' \cite{Cal03b}.  From a historical
perspective, however, it is probably too soon to assess the extent to which
anyone can truly be said to have anticipated the detection of dark energy by
the supernova teams in 1998.  There are similarities here to the detection
of cosmic microwave background (CMB) in 1965, perhaps the only event of
comparable significance in cosmology since the discovery of cosmic expansion
itself, whose ``pre-discovery'' by numerous people is still a subject of
historical controversy after forty years \cite{Kra96c}.  In the case
of dark energy, astronomers had been led to closed, BN-P-like models with
high ratios of $\Olam$ to $\Omat$ on at least three previous occasions.
These models were never conclusively ruled out; rather, interest in them
simply faded when the observational phenomena which seemed to call for
them became less compelling.  

The common feature that unites models of this kind is the existence of a
quasi-static ``loitering period'' that occurs as a result of the struggle
between the self-gravity of ordinary matter and the gravitational
self-repulsion of vacuum energy.  (Acceleration is inevitable in such
models because the energy density of ordinary matter drops with expansion,
whereas that of vacuum energy remains constant and eventually dominates.)
Georges Lema\^{\i}tre first invoked such a loitering phase in the 1930s in
response to an early variant of the age problem: measurements of Hubble's
constant at that time wrongly suggested that the age of the universe would
otherwise be less than the age of the Earth \cite{Kra96d}.  (He also pointed
out that the quasi-static phase would be ideal for the growth of large-scale
structure, an observation that remains valid today.)  Loitering models
reappeared in the late 1960s as a way to explain the puzzling concentration
of quasar redshifts near $z\sim2$ [135-137]. 
Most cosmologists have since preferred to attribute this phenomenon to
evolutionary and/or astrophysical factors, though a definitive explanation
is still lacking.  Again, it was pointed out in the early 1990s
that the rapid growth of structure during a loitering phase could reconcile
data on galaxy clustering with the lack of anisotropy then seen in the CMB
\cite{Sah92,Fel93}; interest in this proposal waned when the hoped-for
anisotropies were finally detected by COBE.  Irrespective of whether it is
eventually seen as something more, the BN-P model certainly holds a place
in this respectable tradition.

The main objection that was raised against loitering models during their
later historical revivals goes back to Robert Dicke in the 1960s and has
been argued most forcefully by Peebles \cite{Pee93}.  The Dicke ``coincidence
argument,'' as it is known, is essentially the observation that any
quasi-static phase long enough to be interesting (relative to the Hubble
timescale) requires a degree of fine-tuning in the cosmological parameters
$\Olam$ and $\Omat$.  As we have seen, however, by 1998 such arguments were no
longer persuasive enough to overcome the overpowering observational case for a
large $\Lambda$ term.  Indeed, cosmologists have now been obliged to accept
values for $\Olam$ and $\Omat$ that are widely regarded as ``preposterous''
by virtue of their nearness to each other \cite{Car01}.  This ``coincidence
problem'' itself pales into insignificance beside the grand-daddy of all
conundrums, the cosmological-constant problem, which implies that all
contributions to the vacuum-energy density cancel each other out to a
precision of some 120 decimal places --- while agreeing with observation at
the 121st and 122nd.  Until such an absurdity is better understood, questions
of fine-tuning to order of magnitude are, to a certain extent, like worries
about whether or not one is whistling off-key in the middle of a hurricane.

For the time being, cosmological models are best judged on their empirical
merits alone.  The measurements of Wolfgang Priester and his colleagues in
1992 using quasar spectra yielded significantly larger values of $\Olam$
(and smaller values of $\Omat$) than those reported six years later using
supernovae.  The BN-P universe is not only dominated by dark energy; it
contains almost {\em nothing else\/}, putting it perilously close to upper
limits on $\Olam$ based on other phenomena such as the statistical frequency
and maximum observed redshift of gravitational lenses.  Its low matter density
leaves little or no room for cold dark matter, which (although it has not been
detected directly) is thought by many to have been necessary as a sort of
``cosmological midwife'' during the birth of large-scale structure at early
times.  Also, its lengthy loitering phase implies a universe considerably
older than anything yet seen in it, unless the Hubble parameter lies at the
upper end of experimental limits --- a ``reverse age problem'' which, while
not necessarily fatal, does raise the question of why we should find ourselves
in an unusually young corner of the cosmos.  On the other hand, the model's
long gestation period eases problems with conventional structure formation
theory, and some support for the high ratio of $\Olam$ to $\Omat$ appears
in the unexpectedly low amplitude of the second peak in the CMB power
spectrum.

The study of these issues occupied most of the last ten years of Wolf's
career, including two more review articles in {\it Naturwissenschaften},
a chapter in the authoritative {\it Bergmann-Schaefer Lehrbuch der
Experimentalphysik} and many more publications [94-109]. 
We will not go over this material in detail here; recent reviews may be found
in \cite{wolf01} and \cite{Tho01}.  Instead, we will approach the phenomenon
of quasar absorption lines in a different way and focus on what is probably
the ``Achilles heel'' of the BN-P approach: its assumption that intrinsic
evolution in the absorber population can be neglected relative to the Hubble
expansion rate.  A similar assumption underlies measurements based on
supernovae, of course, but in that case there is a wider consensus that
systematic evolutionary effects are unlikely to be important.  In the case of
\Lya\ absorbers, no such consensus has emerged, and we will argue that a
generalization of the BN-P method that allowed for intrinsic evolution might
have yielded a more widely accepted detection of dark energy and cosmic
acceleration.

\section{Evolution and the Lyman-$\alpha$ forest} \label{lyalpha}

A general expression for absorption-line number density follows directly from
the space density of absorbers, which may be expressed in a standard isotropic
and homogeneous universe as
\beq
n(z) \propto (1+z)^{3+\eta} \; .
\label{etaDefn}
\eeq
Here $\eta$ parametrizes intrinsic evolution in the absorber population.
If $\eta=0$, then comoving number density is constant and density goes as
$(1+z)^3$, as appropriate for a pressureless fluid of conserved particles.
The line-number density per redshift increment is the derivative of
Eq.~(\ref{etaDefn}) with regard to $z$:
\beq
\frac{dn}{dz} \propto n(z)\frac{R}{\sqrt{1-kr^2}}\frac{dr}{dz} \propto n(z)\frac{dt}{dz} \; .
\label{TheoryRaw}
\eeq
Since $dt/dz = -[(1+z)H(z)]^{-1}$ where Hubble's parameter
$H(z)=H_0[\Omat (1+z)^3 + \Olam - (\Otot-1)(1+z)^2]^{1/2}$ and $H_0$ is
Hubble's constant, Eq.~(\ref{TheoryRaw}) can be expressed in terms of the
total matter density $\Omat$ and the dark-energy density $\Olam$ as
\beq
\frac{dn}{dz} = \frac{\zeta (1+z)^{2+\eta}}{\sqrt{\Omat (1+z)^3 + \Olam - 
   (\Otot - 1)(1+z)^2}} \; .
\label{Theory}
\eeq
Here $\Otot=\Omat+\Olam$ and $\zeta$ is a constant.

If the cosmological constant $\Lambda$ is {\em zero\/}, as was once widely
thought, then we can follow standard texts \cite{Pad93,Pea99} and rewrite
Eq.~(\ref{Theory}) in the form
\beq
\frac{dn}{dz} \propto \frac{(1+z)^{1+\eta}}{\sqrt{1+\Omat z}} \; 
   \hspace{1cm} \mbox{($\Lambda=0$)} \; .
\eeq
Thus for example in the Einstein-de~Sitter (EdS) model with $\Olam=0$ and
$\Omat=1$:
\beq
\frac{dn}{dz} \propto (1+z)^{\eta+1/2} \;
   \hspace{1cm} \mbox{(EdS)} \; .
\label{TheoryEdS}
\eeq
To compare this with observation, a recent analysis of eight high-resolution,
high signal-to-noise quasar spectra using VLT/UVES leads to the following
best-fit expression over $1.5\leqslant z\leqslant 4$ \cite{Kim02}:
\beq
\frac{dn}{dz} = 6.1 (1+z)^{2.47\pm0.18} \; .
\label{Obsn}
\eeq
Eqs.~(\ref{TheoryEdS}) and (\ref{Obsn}) together imply that in an EdS
universe, the number density and/or absorption cross-section of Ly$\alpha$
absorbers must climb very steeply with redshift, $\eta\approx2$.  This is the
basis for the traditional argument that intrinsic evolution is likely to
frustrate any attempt to extract useful limits on $\Omat$ and $\Olam$ from
observations of the \Lya\ forest \cite{Car92}.

In light of the fact that dark energy is now known to exist, however, this
argument must be re-examined.  Is strong intrinsic evolution still required
to fit the data in a $\Lambda$-dominated universe?

\begin{figure}
\resizebox{\hsize}{!}
{\includegraphics{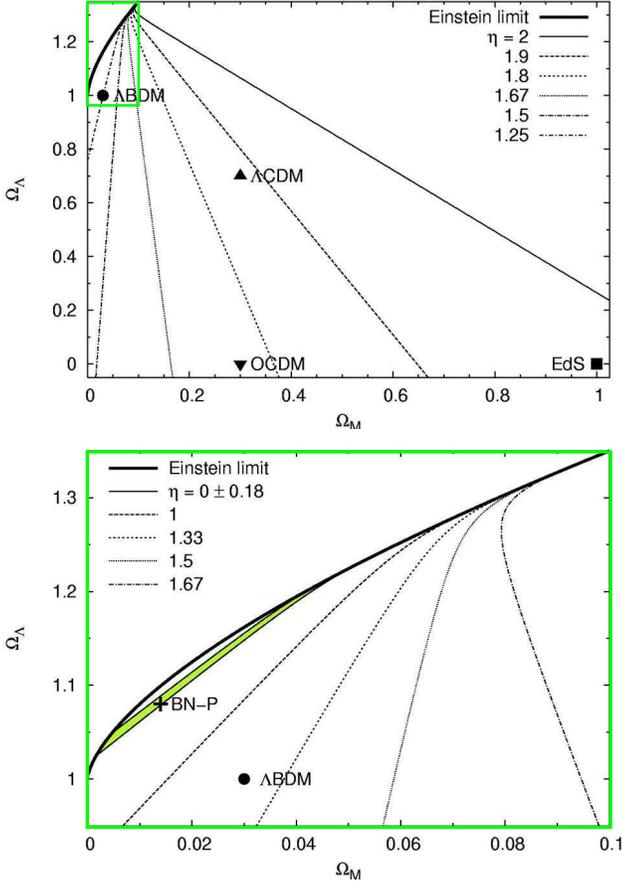}}
\caption{{\bf(a)~Top:} best-fit values of the parameter $\eta$ governing
   the evolution of the Ly$\alpha$ forest according to Eq.~(\ref{etaDefn})
   in the phase space defined by $\Omat$ and $\Olam$.  World models above
   the solid curve marked ``Einstein limit'' are of the ``big bounce''
   rather than ``big bang'' type; their scale size drops to a nonzero
   minimum and then begins to climb again in the past direction.
   {\bf(b)~Bottom:} enlargement of the rectangular region at the upper left
   corner of Fig.~\ref{fig5}(a).  Inside the shaded region, the observational
   data on $dn/dz$ are satisfied with no evolution in the Ly$\alpha$ forest;
   i.e., $\beta=2.47\pm0.18$ and $\eta=0$.  The labelled points correspond
   to models discussed in the main text: EdS ($\Omat=1,\Olam=0$), OCDM
   (0.3,0), $\Lambda$CDM (0.3,0.7), $\Lambda$BDM (0.03,1) and BN-P
   (0.014,1.08).}
\label{fig5}
\end{figure}

The problem can be posed mathematically as follows: for any given model in
the wider cosmological phase space defined by ($\Omat,\Olam$), what values
of $\eta$ best fit the theoretical prediction of Eq.~(\ref{Theory}) to the
observational data in Eq.~(\ref{Obsn})?  And in particular, is any part of
this phase space compatible with $\eta\approx 0$?  We rewrite
Eqs.~(\ref{Theory}) and (\ref{Obsn}) for convenience as
\beqa
\fth(x) & = & \frac{\zeta x^{2+\eta}}{\sqrt{\Omat x^3 + \Olam - 
   (\Omat + \Olam - 1)\,x^2}}
   \label{fTh} \\
\fobs(x) & = & \alpha x^{\beta} \; ,
   \label{fObs}
\eeqa
where $x\equiv (1+z)$.  We are interested in the {\em slopes\/} of $\fth(x)$
and $\fobs(x)$; hence the relevant dimensionless functions are:
\beqa
\Fth(x) & \equiv & \frac{\fth^{\prime}(x)}{\fth(x)} = \frac{2+\eta}{x} 
\nonumber \\
   & & - \frac{[3\Omat x^2 - 2 (\Otot-1)\,x^2]}{2[\Omat x^3 +
   \Olam - (\Otot-1)\,x^2]}  \\
\Fobs(x) & \equiv & \frac{\fobs^{\prime}(x)}{\fobs(x)} = \frac{\beta}{x} \; .
\eeqa
We seek the value of $\eta$ that minimizes the square of the difference
between $\Fth(x)$ and $\Fobs(x)$, averaged over $2.5\leqslant x\leqslant 5$.
In other words,
\beq
\frac{d}{d\eta} \left\{ \int_{x_1}^{x_2} \left[ \Fth(x)-\Fobs(x)
   \right]^2 dx \right\} = 0 \; .
\label{leastSquare}
\eeq
The required best-fit value of $\eta$ is given by
\beqa
\eta & = & \beta - 2 + \left( \frac{x_1 x_2}{x_2 - x_1} \right) \nonumber \\
     & & \times \int_{x_1}^{x_2}
         \frac{\left[ 3\Omat x - 2 (\Otot - 1) \right] \, dx}
              {2\left[ \Omat x^3 + \Olam - (\Otot - 1)\,x^2 \right] } \; .
\label{bestfit}
\eeqa
Eq.~(\ref{bestfit}) is plotted in Fig.~\ref{fig5} in the form of
``iso-evolution'' contours $\eta(\Omat,\Olam)=$const., with $x_1=2.5$,
$x_2=5$ and $\beta=2.47$ as suggested by Eq.~(\ref{Obsn}).

Fig.~\ref{fig5}(a) confirms that strong evolution is required across most
of the phase space defined by $\Omat$ and $\Olam$, including the EdS model
($\eta\approx 2$), the $\Lambda$CDM model with $\Omat=0.3$ and $\Olam=0.7$
($\eta\approx 1.9$) and the open cold dark matter or OCDM model with
$\Omat=0.3$ and $\Olam=0$ ($\eta\approx 1.8$).  Also shown for comparison
is a $\Lambda+$baryonic dark matter or $\Lambda$BDM model with $\Omat=0.03$
(a lower limit on the matter density, corresponding to current estimates for
the baryonic matter density $\Obar$) and $\Olam=1$ (a typical upper limit
from, e.g., gravitational lensing statistics).  This model still requires
moderate evolution with $\eta=5/4$.

Fig.~\ref{fig5}(b) is a close-up of the rectangle at the upper left-hand
corner of Fig.~\ref{fig5}(a).  Of special interest is the shaded region
bounded by thin solid lines.  Within this region, Eq.~(\ref{Theory}) provides
a best fit to Eq.~(\ref{Obsn}) with no evolution at all; i.e. $\eta=0\pm0.18$.
The BN-P model lies almost precisely in the center of this ``no-evolution''
region, thus confirming the conclusion of Wolfgang Priester and his colleagues
that closed and heavily vacuum-dominated models are singled out by the data on
quasar absorption lines over this redshift range --- {\em provided that\/}
$\eta=0$.  The level of agreement is actually striking, because we have used
newer experimental data and a different method of analysis.  While
unsophisticated, the approach followed here has the merit of showing very
quickly and unambiguously that the BN-P results stand or fall on the validity
of the underlying assumption that intrinsic evolution in the \Lya forest can
be neglected.  It also suggests that a modification of the BN-P approach to
take evolution into account might lead to values of $\Omat$ and $\Olam$
in better agreement with those obtained six years later using Type~Ia
supernovae.

\begin{figure}
\resizebox{\hsize}{!}
{\includegraphics{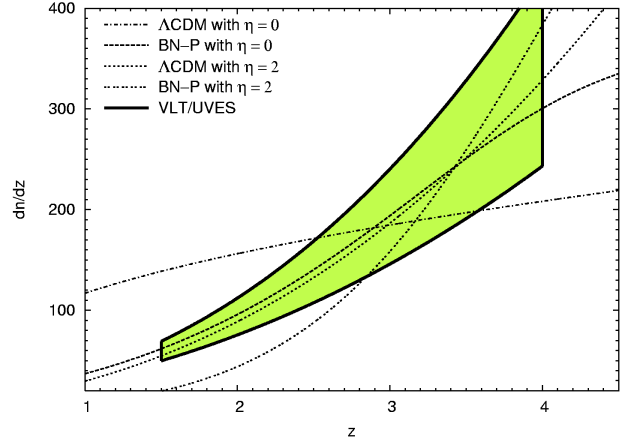}}
\caption{Least-squares fits of Eq.~(\ref{Theory}) to the data on \Lya line
   number density, Eq.~(\ref{Obsn}), for the $\Lambda$CDM and BN-P models
   assuming either no evolution ($\eta=0$) or strong evolution ($\eta=2$)
   in the \Lya forest.  The $\Lambda$CDM model requires strong evolution,
   while the BN-P model is consistent with none.}
\label{fig6}
\end{figure}

As a check on the above results, Fig.~\ref{fig6} compares the predictions of
Eq.~(\ref{Theory}) to the data in Eq.~(\ref{Obsn}) using best-fit values of
$\zeta$ for each model over $1.5\leqslant z\leqslant 4$.
(The EdS, OCDM and $\Lambda$BDM curves have been omitted for clarity; they
are close to $\Lambda$CDM.)  Fig.~\ref{fig6} confirms that the data on $n(z)$
can be fit either by a $\Lambda$CDM model with strong evolution
($\eta\approx2$), or by the BN-P model with no evolution at all ($\eta=0$).

There are several possible reasons why the comoving number density of
absorbers might evolve with redshift.  Numerical simulations suggest ---
under the assumption of large quantities of cold dark matter --- that the
intergalactic medium (IGM) which is responsible for \Lya absorption lines
passes through several distinct ``topological'' stages \cite{Rau98}, as
depicted schematically in Fig.~\ref{fig4}.  At the highest redshifts, this
material is distributed in the form of two-dimensional bubble walls
around overdense regions.  Later on (at intermediate redshifts) the IGM
condenses to form one-dimensional filaments which roughly trace the boundaries
of the overdense regions.  And at the lowest redshifts, these filaments in
turn collapse to zero-dimensional knots.  In such a picture, the slope of the
absorption-line number density should bend upward at intermediate and high
redshifts since lines of sight to distant quasars will ``miss'' much of the
absorbing material in knots at low redshifts, while intercepting most of it
in bubble walls at high redshifts.  But while this picture is qualitatively
easy to grasp, the quantitative implications for $dn/dz$ are not well
established.  Kim et al. (2002) find evidence for one such sharp upturn
at lower redshifts, $z\sim 1$.

Evolution of the Ly$\alpha$ forest will also be driven by changes in the
mean ionization of the IGM.  It has been suggested that neutral hydrogen
column densities in the IGM should decrease with time relative to the ionizing
flux from distant quasars, which remains roughly constant as the Universe
expands \cite{Sto04}.  This would also lead to an increase in \Lya
absorption-line number density with redshift.

Whether these effects are enough to produce the steep evolution that is
required in the context of models such as $\Lambda$CDM remains an open
question.  A generalization of the BN-P approach designed to take gravitational
clustering and changes in ionization into account would be of interest as an
independent probe of the cosmological parameters $\Omat$ and $\Olam$.
If Wolfgang Priester was right, then part of what has been taken as evidence
for rapid evolution may instead be pointing to a closed universe that is even
more strongly vacuum-dominated than has usually been thought.  

\section{Wolf and his legacy}

The breadth of Wolf's interests is reflected by the range of contributions to a
special issue of {\it Naturwissenschaften\/} that was dedicated to him on the
occasion of his sixty-fifth birthday in 1989 by many of those he had gathered
around him in Bonn: Wolfgang Kundt, Peter Blum, Max R\"omer,
Gerd Pr\"olss, J\"org Pfleiderer, Hans Volland, Hans-Joachim Blome,
Hans-J\"org Fahr and Thomas Schmutzler \cite{Kun89}.  Wolf remained
extraordinarily active after his official retirement that year, delivering
his last lectures at the Winter School on Cosmic Evolution in Bad Honnef in
January 2005.  He was indefatigable as a scientist and charming as a host.
He was stubborn in defending his convictions, but always did so with a twinkle
in his eye.  Sooner or later, those who worked with him always heard the
motto, ``If you believe in something, stick your neck out!''
Wolf was well aware of the evidence both for and against his positions,
and kept eager track of the latest experimental data on, for example,
Hubble's constant, the value of which became the subject of a long-running
wager with other members of his Institute.  He carried with him the
culture and erudition of a vanishing age, and delighted in embellishing
his talks and articles with Latin, Greek and even Hebrew flourishes, as well
as the clear and intricate diagrams that were so characteristic of the way
he thought.  Once, when urged by him to assume a value of $H_0$ that seemed
unjustified by observation, one of us went to some trouble to dig up the
perfect quote from Goethe: ``Die Botschaft h\"or ich wohl, allein mir fehlt
der Glaube'' (``The message I hear well; 'tis faith alone I lack'').
Without missing a beat, Wolf responded at once with the line that follows:
``Das Wunder ist des Glaubens liebstes Kind!''  (``The miraculous is faith's
dearest child!'')

\begin{figure}
\resizebox{\hsize}{!}
{\includegraphics{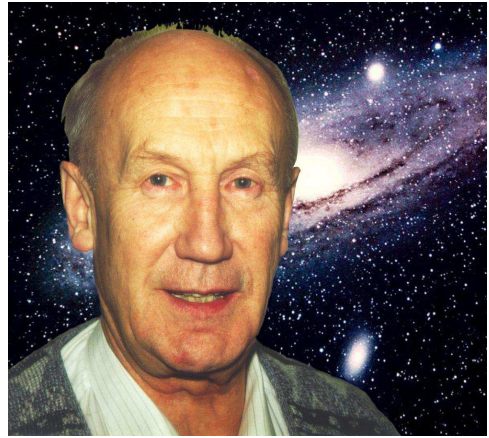}}
\caption{Wolfgang Priester with an image of the Andromeda galaxy (image
   courtesy Hans-J\"org Fahr, 2004).}
\end{figure}

As a cosmologist, Wolf would have wanted to be remembered for the three key
tenets of his preferred world model: a closed universe ($k=+1$) dominated by
vacuum energy without ``exotic'' cold dark matter ($\Olam\gtrsim1$,
$\Omat\sim\Obar$) and with a high value for Hubble's constant
($H_0\gtrsim80$~km~s$^{-1}$~Mpc$^{-1}$).
Of these, the jury is still out on the first, while the second and third
appear increasingly at odds with cosmological consensus.  However,
Wolfgang Priester's real legacy does not lie in the details of the model
he espoused, but in the way he inspired those who were fortunate enough
to know him.  In bidding farewell to a teacher, colleague and dear friend,
we can do no better than borrow from Albert Einstein, who wrote in 1918:
``The supreme task of the physicist is to arrive at those universal
elementary laws from which the cosmos can be built up by pure deduction.
There is no logical path to these laws; only intuition, resting on
sympathetic understanding, can lead to them ... The state of mind that
enables a man to do work of this kind is akin to that of the religious
worshipper or the lover; the daily effort comes from no deliberate intention
or program, but straight from the heart.''

\section*{Acknowledgments}

We are indebted to Wolf's longtime coworker Wolfgang Kundt for his friendly
comments and helpful recollections in Ref.~\cite{Kun06}.


\begin{thebibliography}{199}
\bibitem{wolf53} W. Priester, 
   ``Photometrie von Fraunhofer-Linien mit der Lummer-Platte, angewandt auf
   die Mitte-Rand-Variation der Natrium D-Linien,''
   {\it Zeits. Astrophys.} {\bf 32} (1953) 200-250
\bibitem{wolf57a} W. Priester, 
   ``Photometrie von Fraunhofer-Linien mit der Lummer-Platte, angewandt auf
   die Mitte-Rand-Variation der Natrium D-Linien,''
   {\it Ver\"offentlichungen der Universit\"ats-Sternwarte zu G\"ottingen}
   {\bf 6} (1957) 85-136
\bibitem{wolf58a} W. Priester, 
   ``Photometrie von Fraunhofer-Linien mit der Lummer-Platte, angewandt auf
   die Mitte-Rand-Variation der Natrium D-Linien,''
   {\it Ver\"offentlichungen der Universit\"ats-Sternwarte zu G\"ottingen}
   {\bf 6} (1958) 136.1-136.2
\bibitem{wolf54a} W. Priester, 
   ``Zur Deutung der extragalaktischen Radiofrequenz-Strahlung,''
   {\it Zeits. Astrophys.} {\bf 34} (1954) 283-294
\bibitem{wolf54b} W. Priester, 
   ``\"Uber die Anzahl der Radio-Sterne in der Milchstra{\ss}e,''
   {\it Zeits. Astrophys.} {\bf 34} (1954) 295-301
\bibitem{wolf55a} W. Priester, 
   ``Gest\"orte Multipletts in Sternatmosph\"aren,''
   {\it Zeits. Astrophys.} {\bf 36} (1955) 230-239
\bibitem{wolf55b} W. Priester and F. Dr\"oge, 
   ``\"Uber die Mitte-Rand-Variation der solaren Radiofrequenzstrahlung von
   198 MHz w\"ahrend der Finsternis 1954 Juni 30,''
   {\it Zeits. Astrophys.} {\bf 37} (1955) 132-142
\bibitem{wolf55c} W. Priester, 
   ``\"Uber die Radioquelle Sagittarius A,''
   {\it Zeits. Astrophys.} {\bf 38} (1955) 73-80
\bibitem{wolf56} F. Dr\"oge and W. Priester, 
   ``Durchmusterung der allgemeinen Radiofrequenz-Strahlung bei 200 MHz,''
   {\it Zeits. Astrophys.} {\bf 40} (1956) 236-248
\bibitem{wolf57b} H.G. M\"uller, W. Priester and G. Fischer, 
   ``Radioemission des Kometen 1956 h,''
   {\it Die Naturwissenschaften} {\bf 44} (1957) 392-393
\bibitem{wolf58b} W. Priester, 
   ``Zur Statistik der Radioquellen in der relativistischen Kosmologie,''
   {\it Zeits. Astrophys.} {\bf 46} (1958) 179-202
\bibitem{wolf59a} B.-H. Grahl and W. Priester, 
   ``Eine Messung der Position der Radioquelle Sagittarius A,''
   {\it Zeits. Astrophys.} {\bf 47} (1959) 50-53
\bibitem{wolf61a} W. Priester, 
   ``Die galaktische Radiostrahlung,''
   {\it Mitteilungen der Astronomischen Gesellschaft} {\bf 14} (1961) 21~pp
\bibitem{wolf82a} G. Haslam, R. Wielebinski and W. Priester, 
   ``Radio maps of the sky,''
   {\it Sky \& Telescope} {\bf 63} (1982) 230-232
\bibitem{wolf58c} W. Priester, H.-G. Bennewitz and P. Lengr\"usser, 
   ``Radiobeobachtungen des ersten k\"unstlichen Erdsatelliten,''
   {\it Wissenschaftliche Abhandlungen der Arbeitsgemeinschaft f\"ur
   Forschung des Landes Nordrhein-Westfalen} {\bf 1} (1958) 46~pp
\bibitem{wolf58e} W. Priester 
   and G. Hergenhahn, 
   ``Bahnbestimmung von Erdsatelliten aus Dopplereffecktmessungen,''
   {\it Wissenschaftliche Abhandlungen der Arbeitsgemeinschaft f\"ur
   Forschung des Landes Nordrhein-Westfalen} {\bf 8} (1958) 38~pp
\bibitem{wolf59c} W. Priester, 
   ``Sonnenaktivit\"at und Abbremsung der Erdsatelliten,''
   {\it Mitteilungen der Sternwarte Bonn} {\bf 24} (1959) 4~pp;
   {\it Die Naturwissenschaften} {\bf 46} (1959) 197-198
\bibitem{wolf60a} W. Priester and H.A. Martin, 
   ``Solare und tageszeitliche Effekte in der Hochatmosp\"are aus
   Beobachtungen k\"unstlicher Erdsatelliten,''
   {\it Mitteilungen der Sternwarte Bonn} {\bf 29} (1960) 53~pp;
   {\it Forschungsbericht des Landes Nordrhein-Westfalen} {\bf 547}
   (K\"oln-Opladen: Westdeutscher Verlag, 1960) 53~pp 
\bibitem{wolf60b} H.A. Martin and W. Priester, 
   ``Measurement of solar and diurnal effects in the high atmosphere by
   artificial satellites,'' {\it Nature} {\bf 185} (1960) 600-601
\bibitem{wolf60c} W. Priester and H.A. Martin, 
   ``Temperature inversion in the F1-layer,''
   {\it Nature} {\bf 188} (1960) 200-202
\bibitem{wolf60d} W. Priester, H.A. Martin and K. Kramp, 
   ``Diurnal and seasonal density variations in the upper atmosphere,''
   {\it Nature} {\bf 188} (1960) 202-204
\bibitem{wolf61c} W. Priester, 
   ``Solar activity effect and diurnal variation in the upper atmosphere,''
   {\it J. Geophys. Res.} {\bf 66} (1961) 4143-4148
\bibitem{wolf61d} H.A. Martin, W. Neveling, W. Priester 
   and M. Roemer,
   ``Model of the upper atmosphere from 130 through 1600~km derived from
   satellite orbits,'' {\it Mitteilungen der Sternwarte Bonn} {\bf 35} (1961)
   16~pp; in H.C. van de Hulst, C. de Jager and A.F. Moore (eds),
   {\it Space Research II}, proc. Second International Space Science
   Symposium (Amsterdam: North-Holland, 1961) 902-917
\bibitem{wolf62a} W. Priester and D. Cattani, 
   ``On the semiannual variation of geomagnetic activity and its relation to
   the solar corpuscular radiation,''
   {\it J. Atmospheric Sciences} {\bf 19} (1962) 121-126
\bibitem{wolf63a} W. Priester, M. Roemer and T. Schmidt-Kaler, 
   ``Apparent relation between solar activity and the 440~Mc/s radar
   distance of Venus'' {\it Mitteilungen der Sternwarte Bonn} (1962) 2~pp;
   {\it Nature} {\bf 196} (1963) 464-465
\bibitem{wolf63b} W. Priester (ed), 
   {\it Space Research III}, proc. Third International Space Science Symposium
   (Amsterdam: North-Holland, 1963) 1275~pp
\bibitem{wolf63c} W. Priester, 
   ``Discussion of atmospheric heat sources based on the analysis of satellite
   drag data,'' in M. Roy (ed), {\it Dynamics of Satellites}
   (Berlin: Springer-Verlag, 1963) 143-157
\bibitem{wolf64} G. Newton, R. Horowitz and W. Priester, 
   ``Atmospheric densities from {\it Explorer 17} density gages and a
   comparison with satellite drag data,''
   {\it J. Geophys. Res.} {\bf 69} (1964) 4690-4692
\bibitem{wolf65d} G.P. Newton, R. Horowitz 
   and W. Priester, ``Atmospheric density and temperature variations from the
   {\it Explorer XVII} satellite and a further comparison with satellite drag,''
   {\it Planetary and Space Science} {\bf 13} (1965) 599-616
\bibitem{wolf65e} W. Priester, 
   ``On the variations of the thermospheric structure,''
   {\it Proc. R. Soc. London} {\bf A288} (1965) 493-509
\bibitem{wolf67a} W. Priester, M. R\"omer 
   and H. Volland, ``The physical behavior of the upper atmosphere deduced
   from satellite drag data,'' {\it Sp. Sci. Rev.} {\bf 6} (1967) 707-780
\bibitem{wolf67b} W. Priester, 
   ``Density and temperature variations above 150~km,''
   {\it Bull. Am. Meteorological Soc.} {\bf 48} (1967) 215 
\bibitem{wolf69a} H.E. Newell, M.G. Kroshkin 
   and W. Priester, {\it Satelliten erkunden Erde und Mond}
   (Frankfurt a.M.: Umschau-Verlag, 1969) 136~pp
\bibitem{wolf72} H. Volland, C. Wulf-Mathies and W. Priester, 
   ``On the annual and semiannual variations of the thermospheric density,''
   {\it J. Atmospheric and Terrestrial Physics} {\bf 34} (1972) 1053-1063
\bibitem{wolf76} P. Blum, W. Priester, K. Schuchardt and 
   C. Wulf-Mathies, ``On the decay of satellite orbits,'' in
   {\it Space Research XVI}, proc. Open Meetings of Working Groups on Physical
   Sciences and Symposium and Workshop on Results from Coordinated Upper
   Atmosphere Measurement Programs (Berlin: Akademie-Verlag, 1976) 197-201
\bibitem{wolf85a} K.G.H. Schuchardt, W. Priester, 
   P.W. Blum and H.G. Peters, ``Lower thermospheric density structure derived
   from late decay phases of satellite orbits,''
   {\it Advances in Space Research} {\bf 5} (1985) 179-182
\bibitem{wolf86a} K.G.H. Schuchardt, W. Priester, 
   P.W. Blum and H.G. Peters, ``Anomalous perigee shift and eccentricity
   variation due to air drag in the reentry phase,'' in {\it Re-entry of
   Space Debris} (Noordwijk: ESA Publications, 1986) 55-59
\bibitem{wolf62b} I. Harris and W. Priester, 
   ``Time-dependent structure of the upper atmosphere,''
   {\it J. Atmospheric Sciences} {\bf 19} (1962) 286-301
\bibitem{wolf62c} I. Harris and W. Priester, 
   ``Theoretical models for the solar-cycle variation of the upper
   atmosphere,'' {\it J. Geophys. Res.} {\bf 67} (1962) 4585-4591
\bibitem{wolf62d} I. Harris and W. Priester, 
   ``Time-dependent structure of the upper atmosphere,''
   {\it NASA Technical Note} {\bf D-1443} (June 1962) 71~pp
\bibitem{wolf62e} I. Harris and W. Priester, 
   ``Theoretical models for the solar-cycle variation of the upper atmosphere,''
   {\it NASA Technical Note} {\bf D-1444} (June 1962) 261~pp
\bibitem{wolf63d} I. Harris and W. Priester, 
   ``Heating of the upper atmosphere,'' in W.~Priester (ed),
   {\it Space Research III}, proc. Third International Space Science Symposium
   (Amsterdam: North-Holland, 1963) 53-75
\bibitem{wolf63e} I. Harris and W. Priester, 
   ``Relation between theoretical and observational models of the upper
   atmosphere,'' {\it J. Geophys. Res.} {\bf 68} (1963) 5891-5894
\bibitem{wolf65a} I. Harris and W. Priester, 
   ``On the diurnal variation of the upper atmosphere,'' in
   D.G. King-Hele, P. Muller and G. Righini (eds),
   {\it Space Research V}, proc. Fifth International Space Science Symposium
   (Amsterdam: North-Holland, 1965) 1214 
\bibitem{wolf65c} I. Harris and W. Priester, 
   ``Of the diurnal variation of the upper atmosphere,''
   {\it J. Atmospheric Sciences} {\bf 22} (1965) 3-10
\bibitem{wolf68b} I. Harris and W. Priester, 
   ``The structure of the thermosphere and its variations,'' in R.S. Quiroz
   (ed), {\it Meteorological Investigations of the Upper Atmosphere}, proc.
   American Meteorological Society Symposium on Meteorological Investigations
   above 70 Kilometers, {\it Meteorological Monographs} {\bf 9} (1968) 72-81
\bibitem{wolf69b} I. Harris and W. Priester, 
   ``On the semiannual variation of the upper atmosphere,''
   {\it J. Atmospheric Sciences} {\bf 26} (1969) 233-240
\bibitem{wolf65b} W. Priester and J. Rosenberg, 
   ``Extragalactic radio sources,''
   {\it NASA Technical Note} {\bf D-2888} (July 1965);
   in W.N. Hess (ed), {\it Introduction to Space Science}, 1 ed
   (New York: Gordon and Breach, 1965) 823-862
\bibitem{wolf68a} W. Priester and J. Rosenberg, 
   ``Extragalactic radio sources,'' in W.N. Hess and G.D. Wilmot (eds),
   {\it Introduction to Space Science}, 2 ed
   (New York: Gordon and Breach, 1968) 937-981
\bibitem{wolf66} J. Pfleiderer and W. Priester, 
   ``Neuere Ergebnisse der Erforschung der Quasare, Sterne und Weltraum,''
   {\it Sterne und Weltraum} {\bf 5} (1966) 200-205
\bibitem{wolf68c} M. Grewing, J. Pfleiderer and W. Priester, 
   ``Nichtthermische kosmische Strahlungsquellen''
   {\it Forschungsbericht des Landes Nordrhein-Westfalen}
   {\bf 176} (K\"oln-Opladen: Westdeutscher Verlag, 1968) 48~pp
\bibitem{wolf70b} W. Priester, 
   ``Neue Art energiereicher Objekte im Weltraum,''
   {\it Erdol und Kohle Erdgas Petrochemie Vereinigt mit Brennstoff-Chemie}
   {\bf 23} (1970) 702
\bibitem{wolf73a} J. Pfleiderer, W. Priester 
   and W. K\"ohnlein, ``Processes of continuous radio emission,''
   in A. Bruzek and H. Pilkuhn (eds), {\it Lectures on Space Physics 2.
   Sun and Interplanetary Medium, Relativistic Astrophysics}
   (D\"usseldorf: Bertelsmann Universit\"atsverlag, 1973) 127-193
\bibitem{wolf73b} M. Grewing and W. Priester, 
   ``Nichtthermische Strahlungsquellen im Radiofrequenzbereich (Radiogalaxien,
   Quasare, Pulsare),'' {\it Physik und Didaktik} {\bf 3} (1973) 212-225
\bibitem{wolf77} W. Priester, 
   ``Energiereiche Objekte im Kosmos,''
   {\it Stahl und Eisen} {\bf 97} (1977) 1263-1270
\bibitem{wolf78a} W. Priester, 
   ``Astronomie und \"Offentlichkeit,''
   {\it Mitteilungen der Astronomischen Gesellschaft} {\bf 43} (1978) 11-19
\bibitem{wolf78b} W. Priester, 
   ``Fortschritt aus dem Unerwarteten,''
   {\it Sterne und Weltraum} {\bf 17} (1978) 316-318;
   {\it Mitteilungen der Astronomischen Gesellschaft} {\bf 45} (1979) 9-15
\bibitem{wolf80} W. Priester, 
   ``Quasare, Blasare, Schwerkraftstrudel,''
   {\it Phys. Bl.} {\bf 36} (1980) 241-245
\bibitem{wolf81} W. Priester, 
   ``Die strahlungsst\"arksten Objekte am Rande des Universums: die Quasare,''
   {\it Universitas} (Stuttgart: Wissenschaftliche Verlagsgesellschaft
   Stuttgart) {\bf 1981} (1981) 293-304 
\bibitem{wolf82b} W. Priester, 
   ``Neue Fortschritte in der Kosmologie,''
   {\it Universitas} (Stuttgart: Wissenschaftliche Verlagsgesellschaft
   Stuttgart) {\bf 1982} (1982) 827-832 
\bibitem{wolf83a} W. Priester, 
   ``Vom Urknall bis zu schwarzen L\"ochern,''
   {\it Technische Mitteilungen} (Essen: Vulkan Verlag) {\bf 1} (1983) 2-8
\bibitem{wolf83b} W. Priester, 
   ``Wo blieb die Antimaterie?''
   {\it Naturwissenschaftliche Rundschau} {\bf 36} (1983) 11-15
\bibitem{wolf84a} W. Priester, 
   ``Urknall und Evolution des Kosmos -- Fortschritte in der Kosmologie,''
   {\it Nordrhein-Westf\"alische Akademie der Wissenschaften} {\bf 333}
   (Opladen: Westdeutscher Verlag, 1984) 85~pp
\bibitem{wolf84b} H.-J. Blome and W. Priester, 
   ``Urknall und Evolution des Kosmos I. Einstein-Friedmann-Kosmos und das
   Neutrino-Problem,''
   {\it Naturwissenschaften} {\bf 71} (1984) 456-467
\bibitem{wolf84c} H.-J. Blome and W. Priester, 
   ``Urknall und Evolution des Kosmos II. Inflation\"ar modifizierter Urknall
   und Eschatologie der Kosmos,''
   {\it Naturwissenschaften} {\bf 71} (1984) 515-527
\bibitem{wolf84d} H.-J. Blome and W. Priester, 
   ``Vacuum energy in a Friedmann-Lema\^{\i}tre cosmos,''
   {\it Naturwissenschaften} {\bf 71} (1984) 528-531
\bibitem{wolf85b} H.-J. Blome and W. Priester, 
   ``Vacuum energy in cosmic dynamics,''
   {\it Astrophys. Sp. Sci.} {\bf 117} (1985) 327-335
\bibitem{wolf85c} W. Priester, 
   ``Neutrinos and the fate of our Universe,''
   {\it Universitas} (Stuttgart: Wissenschaftliche Verlagsgesellschaft
   Stuttgart) {\bf 1985} (1985) 143-149 
\bibitem{wolf85d} W. Priester, 
   ``Josef Samuelowitsch Shklovsky,''
   {\it Sterne und Weltraum} {\bf 24} (1985) 427 
\bibitem{wolf86b} W. Priester, 
   ``Vom Ursprung des Universums,'' in H. Maier-Leibnitz (ed),
   {\it Zeugen des Wissens} (Mainz: Hase \& K\"ohler, 1986) 127-156
\bibitem{wolf87a} W. Priester and R. Schaaf, 
   ``Carl Wirtz und die Flucht der Spiralnebel,''
   {\it Sterne und Weltraum} {\bf 26} (1987) 376-377
\bibitem{wolf87b} W. Priester, 
   ``Relationship between redshift and recession velocities in an isotropic
   universe,'' {\it Naturwissenschaften} {\bf 74} (1987) 601-602
\bibitem{wolf87c} W. Priester and H.-J. Blome, 
   ``Zum Problem des Urknalls - `Big Bang' oder `Big Bounce'? I,''
   {\it Sterne und Weltraum} {\bf 26} (1987) 83-89
\bibitem{wolf87d} W. Priester and H.-J. Blome, 
   ``Zum Problem des Urknalls - `Big Bang' oder `Big Bounce'? II,''
   {\it Sterne und Weltraum} {\bf 26} (1987) 140-144
\bibitem{wolf91b} H.J. Blome and W. Priester, 
   ``Big Bounce in the very early universe,''
   {\it Astron. Astrophys.} {\bf 250} (1991) 43-49
\bibitem{wolf88a} Y. Chu, J. Hoell, H.-J. Blome 
   and W. Priester,
   ``On the observational discrimination of Friedmann-Lema\^{\i}tre models,''
   in J. Andouze, M.-C. Pelletan and S. Szalay (eds),
   {\it Large scale structures of the universe}, proc. IAU Symposum No. 130
   (Dordrecht: Kluwer Academic, 1988) 517
\bibitem{wolf88b} Y. Chu, J. Hoell, H.-J. Blome 
   and W. Priester,
   ``The observational discrimination of Friedmann-Lema\^{\i}tre models,''
   {\it Astrophys. Sp. Sci.} {\bf 148} (1988) 119-130
\bibitem{wolf88c} W. Priester, 
   ``Yakov Borisovich Zel'dovich,''
   {\it Sterne und Weltraum} {\bf 27} (1988) 79 
\bibitem{wolf88d} J. Hoell and W. Priester, 
   ``Die Evolutionszeit der Quasare, Sterne und Weltraum,''
   {\it Sterne und Weltraum} {\bf 27} (1988) 412-413
\bibitem{wolf88e} W. Priester, 
   ``The universe of Yakov Zel'dovich,''
   {\it Sky \& Telescope} {\bf 76} (1988) 354
\bibitem{wolf89a} N. Kardashev, H.-J. Blome 
   and W.P. Priester, ``Insular baryonic asymmetry in the universe,''
   {\it Comments Astrophys.} {\bf 13} (1989) 87-101
\bibitem{wolf89b} W. Priester, J. Hoell and H.-J. Blome, 
   ``Das Quantenvakuum und die kosmologische Konstante. Wie alt ist das
   Universum?'' {\it Phys. Bl.} {\bf 45} (1989) 51-56
\bibitem{wolf90a} J. Hoell and W. Priester, 
   ``Voids, Walls und Schweizer K\"ase,''
   {\it Sterne und Weltraum} {\bf 29} (1990) 74-75
\bibitem{wolf90b} J. Hoell and W. Priester, 
   ``Ist die fehlende Masse Illusion?''
   {\it Sterne und Weltraum} {\bf 29} (1990) 638-641
\bibitem{wolf91a} J. Hoell and W. Priester, 
   ``Dark matter and the cosmological constant,''
   {\it Comments Astrophys.} {\bf 15} (1991) 127-136
\bibitem{wolf95e} W. Priester, J. Hoell and H.-J. Blome, 
   ``The scale of the universe: a unit of length,''
   {\it Comments. Astrophys.} {\bf 17} (1995) 327-342
\bibitem{wolf91c} J. Hoell and W. Priester, 
   ``Void-structure in the early universe. Implications for a $\Lambda>0$
   cosmology,'' {\it Astron. Astrophys.} {\bf 251} (1991) L23-L26
\bibitem{wolf92a} D.-E. Liebscher, W. Priester 
   and J. Hoell, ``Lyman-alpha forests and the evolution of the universe,''
    {\it Astron. Gesellschaft Abstract Ser.} {\bf 7} (1992) 60
\bibitem{wolf92b} D.-E. Liebscher, W. Priester 
   and J. Hoell, ``A new method to test the model of the universe,''
   {\it Astron. Astrophys.} {\bf 261} (1992) 377-381
\bibitem{wolf92c} D.-E. Liebscher, W. Priester 
    and J. Hoell, ``Lyman-alpha forests and the evolution of the universe,''
   {\it Astron. Nachr.} {\bf 313} (1992) 265-273
\bibitem{wolf94a} J. Hoell and W. Priester, 
   ``The Lyman $\alpha$ forest and the universal bubble structure,'' in
   W. Wamsteker, M.S. Longair and Y. Kondo (eds),
   {\it Frontiers of Space and Ground-Based Astronomy: The Astrophysics of the
   21st Century} (Dordrecht: Kluwer Academic, 1994) 651-652
\bibitem{wolf94b} J. Hoell and W. Priester, 
   ``Galaxy formation in a Friedmann-Lema\^{\i}tre model,'' in
   G. Hensler, C. Theis and J. Gallagher (eds),
   {\it Panchromatic View of Galaxies -- Their Evolutionary Puzzle}
   (Gif-sur-Yvette: \'Editions Fronti\`eres, 1994) 29-33
\bibitem{wolf94c} J. Hoell, D.-E. Liebscher 
   and W. Priester, ``Confirmation of the Friedmann-Lema\^{\i}tre universe
   by the distribution of the larger absorbing clouds,''
   {\it Astron. Nachr.} {\bf 315} (1994) 89-96
\bibitem{wolf94d} W. Priester, 
   ``Neue Erkenntnisse \"uber Ursprung und Entwicklung des Kosmos,''
   {\it Technische Mitteilungen} (Organ des Hauses der Technik e.V. Essen)
   {\bf 87} (1994) 3-12
\bibitem{wolf95a} J. Hoell and W. Priester, 
   ``The Lyman $\alpha$ forest and the universal bubble structure,'' in
   M. Behara, R. Fritsch and R.G. Lintz (eds), {\it Symposia Gaussiana}
   (Berlin: Walter de Gruyter, 1995) 617-625
\bibitem{wolf95b} H.-J. Blome, W. Priester 
   and J. Hoell, ``New ways in cosmology: I. Friedmann-Lema\^{\i}tre model
   derived from the Lyman alpha forest in quasar spectra,'' in M.M. Shapiro,
   R. Silberberg and J.P. Wefel (eds), {\it Currents in high-energy
   astrophysics} (Dordrecht: Kluwer Academic, 1995) 291-300
\bibitem{wolf95c} H.-J. Blome, W. Priester 
   and J. Hoell, ``New ways in cosmology: II. Alternative models for the
   very early universe,'' in M.M. Shapiro, R. Silberberg and J.P. Wefel (eds),
   {\it Currents in high-energy astrophysics}
   (Dordrecht: Kluwer Academic, 1995) 301-312
\bibitem{wolf95d} D.-E. Liebscher and W. Priester, 
   ``Quasar absorption lines and the parameters of the Friedmann universe,''
   in J. M\"ucket, S. Gottloeber and V. M\"uller (eds),
   {\it Large scale structure in the universe}
   (Singapore: World Scientific, 1995) 273
\bibitem{wolf95f} W. Priester, 
   ``\"Uber den Ursprung des Universums: das Problem der Singularit\"at,''
   {\it Nordrhein-Westf\"alische Akademie der Wissenschaften} {\bf 414}
   (Opladen: Westdeutscher Verlag, 1995) 36~pp
\bibitem{wolf96a} W. Priester, J. Hoell and 
   C. van de Bruck, ``Friedmann-Lema\^{\i}tre model derived from the Lyman
   alpha forest in quasar spectra,'' in V. Trimble and A. Reisenegger (eds),
   {\it Clusters, lensing and the future of the universe}, proc. 
   ASP Conference {\bf 88} (1996) 286-289
\bibitem{wolf96b} C. van de Bruck and W. Priester, 
   ``Quasar pairs testing the bubble wall model,'' in
   V. Trimble and A. Reisenegger (eds), 
   {\it Clusters, lensing and the future of the universe}, proc. 
   ASP Conference {\bf 88} (1996) 290-293
\bibitem{wolf96c} W. Priester, J. Hoell, D.-E. Liebscher 
   and C. van de Bruck, ``Friedmann-Lema\^{\i}tre model derived from the
   Lyman alpha forest in quasar spectra,'' in Y.N. Gnedin, A.A. Grib and
   V.M. Mostepaneko (eds), proc. Third Alexander Friedmann International
   Seminar on Gravitation and Cosmology (St. Petersburg: Friedmann
   Laboratory, 1996) 52-67
\bibitem{wolf97} H.-J. Blome, J. Hoell and W. Priester, 
   ``Kosmologie,'' in {\it Bergmann-Schaefer Lehrbuch der Experimentalphysik}
   {\bf 8}, 1 ed (Berlin: Walter de Gruyter, 1997) 311-427
\bibitem{wolf98a} W. Priester and C. van de Bruck, 
   ``75 Jahre Theorie des expandierenden Kosmos: Friedmann Modelle und der
   `Einstein-Limit','' {\it Naturwissenschaften} {\bf 85} (1998) 524-538
\bibitem{wolf98b} C. van de Bruck, M. Soika and W. Priester, 
   ``Aktuelle Modelle der Kosmologie,''
   {\it Astronomie + Raumfahrt} {\bf 35} (1998) 30-33
\bibitem{wolf99} C. van de Bruck and W. Priester, 
   ``The cosmological constant $\Lambda$, the age of the universe and dark
   matter: clues from the Ly$\alpha$-forest,''
   in H.V. Klapdor-Kleingrothaus and L. Baudis (eds),
   {\it Dark98}, proc. Second International Workshop on Dark Matter
   (Bristol: Institute of Physics Press, 1999) 181-196;
   arXiv preprint astro-ph/9810340
\bibitem{wolf01} J. Overduin and W. Priester, 
   ``Problems of modern cosmology: how dominant is the vacuum?''
   {\it Naturwissenschaften} {\bf 88} (2001) 229-248;
   arXiv preprint astro-ph/0101484
\bibitem{wolf02} H.-J. Blome, J. Hoell and W. Priester, 
   ``Kosmologie,'' in {\it Bergmann-Schaefer Lehrbuch der Experimentalphysik}
   {\bf 8}, 2 ed (Berlin: Walter de Gruyter, 2002) 439-582
\bibitem{wolf04} J. Overduin and W. Priester, 
   ``An accelerating closed Universe,''
   in M.M. Shapiro, T. Stanev and J.P. Wefel (eds), {\it Relativistic 
   Astrophysics and Cosmology}, proc. International School of Cosmic Ray
   Astrophysics -- 13th Course (Singapore: World Scientific, 2004) 3-21
\bibitem{Kra96a} H. Kragh, {\it Cosmology and Controversy}
   (Princeton: Princeton University Press, 1996) p~329
\bibitem{Gol65} T. Gold, ``Summary of after-dinner speech,'' in
   {\it Quasi-Stellar Sources and Gravitational Collapse}
   (Chicago: University of Chicago Press, 1965), p~470
\bibitem{Cal03a} N. Calder,
   {\it Magic Universe: The Oxford Guide to Modern Science}
   (Oxford: Oxford University Press, 2003) p~73
\bibitem{Chi64} H.-Y. Chiu,
   ``Gravitational Collapse,'' in
   {\it Physics Today} {\bf 17} (1964) 21-34
\bibitem{Sch70} S. Chandrasekhar (ed), in M. Schmidt,
   {\it Astrophys. J.} {\bf 162} (1970) 371-379
\bibitem{Cal98} R.R. Caldwell, R. Dave and P.J. Steinhardt,
   ``Cosmological imprint of an energy component with general equation of
   state,'' {\it Phys. Rev. Lett.} {\bf 80} (1998) 1582-1585
\bibitem{Pap71} D.A. Papanastassiou and G.F. Wasserburg,
   ``Lunar Chronology and Evolution from Rb-Sr Studies of Apollo 11 and 12
   Samples,'' {\it Earth and Planetary Science Lett.} {\bf 11} (1971) 37-62
\bibitem{Lew75} R.B. Lewis, B. Srinivasan and E. Anders,
   ``Host Phase of a Strange Xenon Component in Allende,''
   {\it Science} {\bf 190} (1975) 1251-1262
\bibitem{Wet88} C. Wetterich,
   ``Cosmology and the fate of dilatation symmetry,''
   {\it Nucl. Phys.} {\bf B302} (1988) 668-696
\bibitem{Pee88} P.J.E. Peebles and B. Ratra,
   ``Cosmology with a time-variable cosmological ``constant'',''
   {\it Astrophys. J.} {\bf 325} (1988) L17-L20
\bibitem{Str75} E. Streeruwitz,
   ``Vacuum fluctuations of a quantized scalar field in a Robertson-Walker
   universe,'' {\it Phys. Rev.} {\bf D11} (1975) 3378-3383
\bibitem{Isr89} M. Israelit and N. Rosen, ``A singularity-free cosmological
   model in general relativity,'' {\it Astrophys. J.} {\bf 342} (1989) 627-634
\bibitem{Kra96b} H. Kragh, in Ref.~\cite{Kra96a} pp~75, 99, 114 and 208
\bibitem{Pet82} V. Petrosian,
   ``Phase transitions and dynamics of the Universe,''
   {\it Nature} {\bf 298} (1982) 805-808
\bibitem{Gra03} S. Gratton and P. Steinhardt,
   ``Cosmology -- beyond the inflationary border,''
   {\it Nature} {\bf 423} (2003) 817-818
\bibitem{Gas03} M. Gasperini and G. Veneziano,
   ``The pre-big-bang scenario in string cosmology,''
   {\it Phys. Rep.} {\bf 373} (2003) 1-212;
   arXiv preprint hep-th/0207130
\bibitem{Boj05} M. Bojowald,
   ``The early universe in loop quantum gravity,''
   {\it J. Phys. Conf. Ser.} {\bf 24} (2005) 77-86;
   arXiv preprint gr-qc/0503020
\bibitem{Efs90} G. Efstathiou, W.J. Sutherland and S.J. Maddox,
   ``The cosmological constant and cold dark matter,''
   {\it Nature} {\bf 348} (1990) 705-707
\bibitem{OS95} J.P. Ostriker and P.J. Steinhardt,
   ``Cosmic concordance,''
   {\it Nature} {\bf 377} (1995) 600-602
   arXiv preprint astro-ph/9505066
\bibitem{KT95} L.M. Krauss and M.S. Turner,
   ``The cosmological constant is back,''
   {\it Gen. Rel. Grav.} {\bf 27} (1995) 1137-1144
\bibitem{Edd24} A.S. Eddington
   {\it The mathematical theory of relativity}
   (Cambridge: Cambridge University Press, 1924) 154
\bibitem{Cal03b} N. Calder, in Ref.~\cite{Cal03a} p~694
\bibitem{Gla98} J. Glanz,
   ``Breakthrough of the Year: Astronomy -- cosmic motion revealed,''
   {\it Science} {\bf 282} (1998) 2156-2157
\bibitem{Kra96c} H. Kragh, in Ref.~\cite{Kra96a} pp~132-135, 343-347
\bibitem{Kra96d} H. Kragh, {\it ibid\/} pp~52-53
\bibitem{Pet67} V. Petrosian, E. Salpeter and P. Szekeres,
   ``Quasi-stellar objects in universes with non-zero cosmological constant,''
   {\it Astrophys. J.} {\bf 147} (1967) 1222-1226
\bibitem{Shk67} J. Shklovsky,
   ``On the nature of the ``standard'' absorption spectrum of the quasi-stellar
   objects,'' {\it Astrophys. J.} {\bf 150} (1967) L1-L3
\bibitem{Kar67} N. Kardashev,
   ``Lema\^{\i}tre's universe and observations,''
   {\it Astrophys. J.} {\bf 150} (1967) L135-L139
\bibitem{Sah92} V. Sahni, H. Feldman and A. Stebbins,
   ``Loitering universe,'' {\it Astrophys. J.} {\bf 385} (1992) 1-8
\bibitem{Fel93} H.A. Feldman and A.E. Evrard,
   ``Structure in a loitering universe,''
   {\it Int. J. Mod. Phys.} {\bf D2} (1993) 113-122
\bibitem{Pee93} P.J.E. Peebles,
   {\it Principles of Physical Cosmology}
   (Princeton: Princeton University Press, 1993) pp~318-319, 364-367
\bibitem{Car01} S.M. Carroll,
   ``The cosmological constant,''
   {\it Living Rev. Relativity} {\bf 4} (2001) 1-50
\bibitem{Tho01} J. Thomas and H. Schulz,
   ``Incompatibility of a comoving Ly$\alpha$ forest with supernova-Ia
   luminosity distances,''
   {\it Astron. Astrophys.} {\bf 371} (2001) 1-10;
   arXiv preprint astro-ph/0103283
\bibitem{Pad93} T. Padmanabhan, T.,
   {\it Structure Formation in the Universe}
   (Cambridge: Cambridge University Press, 1993),~342
\bibitem{Pea99} J.A. Peacock,
   {\it Cosmological Physics}
   (Cambridge: Cambridge University Press, 1999)~363
\bibitem{Kim02} T.-S. Kim et al.,
   ``The physical properties of the \Lya\ forest at $z>1.5$,''
   {\it Mon. Not. R. Astron. Soc.} {\bf 335} (2002) 555-573;
   arXiv preprint astro-ph/0205237
\bibitem{Car92} S.M. Carroll, W.H. Press and E.L. Turner,
   ``The cosmological constant,''
   {\it Ann. Rev. Astron. Astrophys.} {\bf 30} (1992) 499-542
\bibitem{Rau98} M. Rauch,
   ``The Lyman alpha forest in the spectra of quasistellar objects,''
   {\it Ann. Rev. Astron. Astrophys.} {\bf 36} (1998) 267-316
\bibitem{Sto04} J.T. Stocke, J.M. Shull and S.V. Penton,
   ``The baryon content of the local intergalactic medium,'' in
   {\it From Planets to Cosmology\/} (Baltimore: Space Telescope Science
   Institute, 2005); arXiv preprint astro-ph/0407352
\bibitem{Kun89} W. Kundt (ed), ``Brennpunkte astrophysikalischer Forschung,''
   {\it Naturwissenschaften} {\bf 76} (1989) 289-324
\bibitem{Kun06} W. Kundt, ``Nachruf auf Wolfgang Priester,''
   {\it Telescopium} {\bf 130} (2005) 51-55
\end{thebibliography}
\end{document}